\documentclass[showpacs,twocolumn]{revtex4}
\usepackage{amsmath,amssymb}
\usepackage{url}
\usepackage{graphicx}
\usepackage{epstopdf,epsfig,color}

\begin{document}
\title{Integrating Post-Newtonian Equations on Graphics Processing Units}

\author{Frank Herrmann} 

\affiliation{Department of Physics, Center for Fundamental Physics, and Center for Scientific
  Computation~and Mathematical Modeling, University of Maryland,
  College Park, MD 20742, USA}

\author{John Silberholz} 
\affiliation{Center for Scientific Computation and Mathematical Modeling, University of Maryland,
  College Park, MD 20742, USA}

\author{Mat\'{\i}as Bellone} 
\affiliation{Facultad de Matem\'atica, Astronom\'{\i}a y F\'{\i}sica, Universidad Nacional de C\'ordoba, C\'ordoba 5000, Argentina. }

\author{Gustavo Guerberoff} 
\affiliation{Facultad de Ingenier\'{\i}a, Instituto de Matem\'atica y Estad\'{\i}stica ``Prof. Ing. Rafael Laguardia'',
Universidad de la Rep\'ublica, Montevideo, Uruguay.}

\author{Manuel Tiglio} 

\affiliation{Department of Physics,  Center for Fundamental Physics, and Center for Scientific
  Computation and Mathematical Modeling, University of Maryland,
  College Park, MD 20742, USA}

\begin{abstract}
  We report on early results of a numerical and statistical study of binary black
  hole inspirals. The two black holes are evolved using post-Newtonian
  approximations starting with initially randomly distributed spin
  vectors. We characterize certain aspects of the distribution shortly
  before merger. In particular we note the uniform distribution
  of black hole spin vector dot products shortly before merger and a high
  correlation between the initial and final black hole spin vector dot products
  in the equal-mass, maximally spinning case. These
  simulations were performed on Graphics
  Processing Units, and we demonstrate a speed-up of a factor 50 over a more
  conventional CPU implementation.
\end{abstract}

\pacs{}
\maketitle

\section{Introduction}
With a number of gravitational wave detectors
(LIGO~\cite{Abbott:2007kva}, Virgo~\cite{Acernese:2008zza},
TAMA~\cite{Takahashi:2008zzb}, and GEO600~\cite{Willke:2007zz}) now
measuring signals at design sensitivity, the prospect of direct
detection of gravitational radiation is becoming increasingly real and
hence there is now a definite need to understand as much as possible
about the most likely signal source -- the stellar-mass binary black
hole (BBH) system. Over recent years numerical relativity has reached
the point were accurate and reliable inspiral and merger calculations
can be produced by a number of different
groups~\cite{Aylott:2009tn}. However, these calculations require
significant computational effort, 
and the generation of a large-scale bank of numerical templates
is currently intractable. Recent comparisons between numerical
relativity and post-Newtonian (PN) approximate waveforms show good
agreement surprisingly close to merger (up to a few
orbits)~\cite{boyle-2007,Hannam:2007wf}, corresponding to a
gravitational wave frequency of about $\omega_{\text{GW}}\approx 0.1$
or an orbital frequency of $\omega \approx 0.05$. 

The main purpose of this
work is to start a detailed study of the phase
space of the PN equations of motion (see also~\cite{Schnittman:2004vq}). We randomly choose initial conditions for
those equations corresponding 
to circular orbits at some chosen orbital frequency $\omega_0$. We then 
integrate the PN equations of motion to some termination frequency
$\omega_f$ shortly before merger, but in a region where the 
approximation has still been validated by numerical relativity. There
are a number of interesting physical questions one can address in this
context. It might be that there are certain configurations 
that are preferred shortly before merger. If so, maybe these regions
should be studied using full numerical relativity. Another motivation
regards partial information: If one had from some other observation
(such as an electromagnetic counterpart) enough information about a
BBH system long before merger to estimate a fraction of the
parameters, it would be valuable to be able to estimate the system's
properties shortly before merger. For example, assume that the spin
vector and mass of one BH have been measured, but only the mass of the
other is known. Using an initially uniform distribution for the
unknown spin vector an interesting question would be the final
distribution of spin vectors shortly before merger. Would there be a
uniform distribution or would it be strongly peaked, and how does it
change given changes in the known parameters? We start to address some
of these questions in this work, but much remains to be addressed in
future research.

Another motivation for this work is provided by gravitational
recoil. Numerical relativity has been able to obtain estimates for the
gravitational recoil in
unequal-mass~\cite{Gonzalez:2006md,Baker:2006vn,Herrmann:2006ks} and
spinning~\cite{2007ApJ...661..430H,2007gr.qc.....1163K,2007ApJ...659L...5C}
configurations; in particular, extremely high recoil values
have been obtained for specific
configurations~\cite{2007ApJ...659L...5C,Gonzalez:2007hi,2007gr.qc.....2133C}. Initial
studies using effective-one-body techniques have shown that very large
kicks are unlikely to arise~\cite{2007astro.ph..2641S}. Using the PN
inspiral approach, it might be possible to predict the likelihood of
certain configurations shortly before merger and then relate this
result to the recoil using one of the many recoil formulas available
in the literature.

To perform this study, we use the significant performance advantages
of Graphics Processing Units (GPUs) to study the phase space of the
BBH problem in a post-Newtonian (PN) setting. Each inspiral is
described by a set of ordinary differential equations (ODEs) and the collection of inspirals are all
decoupled. Therefore the computational problem is of the
``embarrassingly parallel'' kind, which is perfectly suited for
GPUs. At this early stage in our studies, in which we focus on small
portions of the full parameter space, the performance benefits
provided by GPUs are not essential to perform the research. However,
we anticipate that once we begin studies of larger sections of the
parameter space the performance gains will become important.

\section{PN Equations of Motion}
We integrate the post-Newtonian equations (Eqs. (1)-(4) and (9)) from
Ref.~\cite{Buonanno:2002fy} (see Erratum~\cite{Buonanno:2002erratum}),
which describe a circular inspiral of 2 spinning BHs. The evolution is
given by a system of coupled ODEs for the orbital frequency $\omega$,
the individual spin vectors $\mathbf{S}_i$ for the 2 BHs, and the unit
orbital angular momentum vector $\hat{\mathbf{L}}_n$.
\begin{widetext}
\begin{eqnarray}
\dot{\omega}&=&\omega^2 \frac{96}{5} \eta (M\omega)^{5/3} \Bigg\{1-\frac{743+924 \eta}{336} (M\omega)^{2/3}\nonumber\\
            & &    -\left(\frac{1}{12} \sum_{i=1,2}\left(\chi_i \hat{\mathbf{L}}_n\cdot\hat{\mathbf{S}}_i (\frac{113 m_i^2}{M^2}+75 \eta)\right)-4 \pi\right) M\omega\nonumber\\
            & &    +\left( \frac{34103}{18144}+\frac{13661}{2016} \eta+\frac{59}{18} \eta^2 \right) (M\omega)^{4/3}-\frac{1}{48} \eta \chi_1 \chi_2 \left(247 (\hat{\mathbf{S}}_1\cdot\hat{\mathbf{S}}_2)-721 (\hat{\mathbf{L}}_n\cdot\hat{\mathbf{S}}_1) (\hat{\mathbf{L}}_n\cdot\hat{\mathbf{S}}_2)\right) (M\omega)^{4/3}\nonumber\\
            & &    -\frac{1}{672} (4159+15876 \eta) \pi (M\omega)^{5/3}+\Bigg(\left(\frac{16447322263}{139708800}-\frac{1712}{105} \gamma_E+\frac{16}{3} \pi^2\right)+(-\frac{273811877}{1088640}+\frac{451}{48} \pi^2-\frac{88}{3} \hat{\theta}\eta) \eta\nonumber\\
            & &    +\frac{541}{896} \eta^2-\frac{5605}{2592} \eta^3-\frac{856}{105} log(16 (M\omega)^{2/3})\Bigg) (M\omega)^2+(-\frac{4415}{4032}+\frac{358675}{6048} \eta+\frac{91495}{1512} \eta^2) \pi (M\omega)^{7/3} \Bigg\}\label{eq:domdt}\\
\dot{\mathbf{S}_i}&=&\mathbf{\Omega}_i \times \mathbf{S}_i\label{eq:dSidt}\\
\dot{\hat{\mathbf{L}}}_n&=&-\frac{(M\omega)^{1/3}}{\eta M^2} \frac{d\mathbf{S}}{dt}\label{eq:dLdt}
\end{eqnarray}
\end{widetext}
where $d\mathbf{S}/dt=d\mathbf{S}_1/dt+d\mathbf{S}_2/dt$,
$\gamma_E=0.577\ldots$ is Euler's constant, and
$\hat{\theta}=1039/4620$. The total mass is denoted by $M=m_1+m_2$ and
$\eta=m_1 m_2/M^2$ is the symmetric mass ratio. The magnitude of the
angular momentum can be computed via $\left|\mathbf{L}_n\right|=\eta M^{5/3}
\omega^{-1/3}$.

The evolution of the individual spin vectors $\mathbf{S}_i$ for the 2
BHs is described by a precession around $\mathbf{\Omega}_i$ with
\begin{widetext}
\begin{equation}
\mathbf{\Omega}_1=\frac{(M\omega)^2}{2M}\left(\eta (M\omega)^{-1/3} (4+3\frac{m_2}{m_1})\hat{\mathbf{L}}_n+1/M^2(\mathbf{S}_2-3(\mathbf{S}_2\cdot\hat{\mathbf{L}}_n)\,\hat{\mathbf{L}}_n)\right)\,,\label{eq:Omega}
\end{equation}
\end{widetext}
and $\mathbf{\Omega}_2$ is obtained by $1\leftrightarrow 2$. Note that
the spin vectors $\mathbf{S}_i$ are related to the spin unit vectors
$\hat{\mathbf{S}}_i$ via $\mathbf{S}_i=\chi_i m_i^2
\hat{\mathbf{S}}_i$, i.e. $\chi_i\in [0,1]$ is the Kerr spin parameter
of BH $i$. The system of coupled ODEs for $\omega, \hat{\mathbf{L}}_n,
$ and $\mathbf{S}_i$ given mass and spin parameters $m_i, \chi_i$ are
integrated from an initial frequency $\omega_0$ to a final frequency
$\omega_f$. Typically, we choose $\omega_0$ corresponding to an initial
separation of $r\approx 40 M$ and $\omega_f=0.05$, which is a
conservative estimate of where the PN equations still
hold~\cite{boyle-2007,Hannam:2007wf}. Integrating the equations in a
range of $\omega$ rather than $t$ provides a slightly more
gauge-invariant measure to compare different systems.

The ODEs are integrated using the Dormand-Prince
method. We set an
error tolerance of $5\times 10^{-7}$ and start with an adaptive time
step of size $h=10$.

The 2 initial spin vectors $\mathbf{S}_i$ span a 6-dimensional
parameter space for each choice of parameters $m_i, \omega_0$, and
$\omega_f$. Note that the BBH problem is scale free, and we can
therefore fix the total mass $M=m_1+m_2=1$ and need to study only the
dependence on $m_1$. The last 2 parameters $\omega_0$ and $\omega_f$
are only interesting as consistency checks, and it is likely
sufficient to analyze the situation for a few select choices. This
still results in a challenging 7-dimensional ($m_1$ \{1 dof\},
$\mathbf{S}_1$ \{3 dof\}, and $\mathbf{S}_1$ \{3 dof\},
dof: degrees of freedom) problem. Here we break down the problem into a much
more manageable 4-dimensional one by fixing not just $m_i$, but
also the spin magnitudes $\chi_i$, which then means only the unit spin
vectors $\hat{\mathbf{S}}_i$ are chosen freely. Each unit spin vector
has only 2 degrees of freedom (since the third is given by the
normalization condition).

\section{Notes on GPU computing}
We use NVIDIA's CUDA runtime and programming system~\cite{cuda} to
implement the PN evolution code on GPUs. The GPUs today provide most
impressive speed-ups over CPUs for single-precision
computations. GPUs are primarily designed for computer games and hence
current and future capabilities are essentially set by that market.
This means it is unlikely that double-precision will be supported
at similar speed-ups as single-precision. While double-precision
arithmetic is now supported on
some of the high-end cards, the performance gains over CPU codes are
not very impressive and in our opinion do not yet justify the effort
of porting the code to the CUDA infrastructure if double-precision is
necessary throughout the code. Note that some results have been
obtained where the code has been examined carefully and only a few
critical computations where done in double-precision, providing most
of the single-precision speed-up. By comparing single-precision and
double-precision CPU results on a number of inspiral configurations we
have verified single-precision is sufficient for our problem.

Porting the code to the CUDA architecture was relatively
straightforward -- we implemented the right-hand-side (RHS)
computations of the ODEs on the GPU in a device kernel. The ODE integrator
runs on the CPU and spawns kernels on the GPU, transferring the initial
state. The full evolution is then performed on the GPU, including all
necessary calculations of the RHS. At the end of the simulation, the
state is transferred back to the CPU for output and analysis.

GPUs demonstrate significant global memory access latency (for the
NVIDIA S1070, 400-600\,cycles~\cite{cuda}), and even worse access times
are encountered for ``non-coalesced'' memory access. The latter is
defined slightly differently on different NVIDIA GPU
generations~\cite{cuda}, but for the latest generation it happens most
frequently if multiple memory segments are accessed within a group of
16 threads called a half-warp. We have measured the non-coalesced memory
stores and loads using the CUDA profiler and found no uncoalesced memory
access, significantly simplifying the programming since this eliminates
the need to avoid certain global memory access patterns using shared
memory.

We found error detection and result
verification on the GPU to be critical exercises. The computations on
the GPUs are surprisingly resilient to errors happening on the
cards. For instance, the on-board memory is not error correcting and
kernel failures are not caught by default. This ``error resilience''
can lead to rather interesting failure modes from a scientific
computing perspective. When too many inspirals are spawned
simultaneously the runtime warns and terminates the program with an
error message. As the number of inspirals is reduced a regime of
``silent'' failure is entered, where the program runs without any
indication of error or warning, but it produces incorrect
results. This can be caught by explicitly checking each kernel, but
the runtime does not generate errors itself. Issues like this
highlight the case for independent error checks for scientific
computing.

We therefore decided to write our code in a multi-threaded way on the
CPU. We first generate the random initial data for our inspiral
studies on the GPU for the maximum number of parallel inspirals that can
run successfully. We then transfer the initial data to the CPU and
select a random subset of typically about 1\% of these inspirals on
the CPU.  While the GPU is performing all the inspirals, we also evolve
the selected inspirals on the CPU in double-precision in a separate
CPU thread. After the GPU is done, we copy the data back to the CPU
memory and compare the single-precision GPU data against the
double-precision CPU data for the selected subset. This also validates
that single-precision does not introduce unacceptable errors in this problem. We believe
such cross checks are currently crucial for GPU computing. We want to
stress however that we have not observed problems or errors once the
initial ones were worked out.

\section{Performance Results}
We now evaluate performance advantages GPUs deliver in this
context. We executed our code on a single core of the quad-core Intel
Xeon E5410 CPU running at 2.33\,GHz, which is rated at around
5\,GFlops in double-precision. Note that the problem is
``embarrassingly parallel,'' so the CPU will be able to provide
excellent scaling over the 4 cores. We integrated one of our test
inspirals in the range $[\omega_0=0.004,\omega_f=0.1]$ 100 times
serially and found we could achieve around 0.057 inspirals per
millisecond (ms) on a single core.

For comparison, one of the four GPUs on our high-end unit (the NVIDIA Tesla
S1070) is rated at 1035\,GFlops in single-precision, delivering a theoretical
performance advantage of about a \emph{factor 200} over the single-core
CPU. We ran our test setup, spawning a large number of
simultaneous inspirals in parallel.
\begin{figure}
\includegraphics[width=0.4\textwidth]{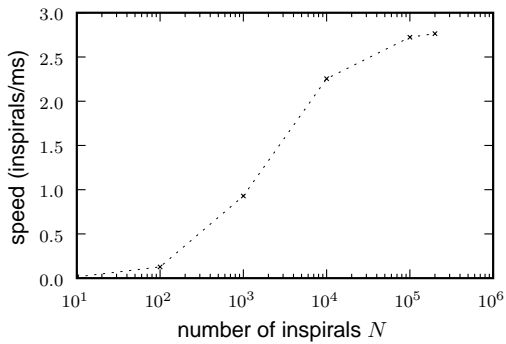}
\caption{Performance of the GPU card. We vary the number of inspirals
  $N$ scheduled simultaneously on the GPU. For fewer than 100,000
  inspirals peak performance is not reached, as there are not enough
  simultaneously executing threads to mask memory access latency. At
  100,000 inspirals saturation is reached as the GPU works serially
  through the extra inspirals.}
\label{fig:perf}
\end{figure}
Fig.~\ref{fig:perf} shows results for the performance of the GPU
card. As we increase the number of simultaneously scheduled inspirals
$N$ the speed levels off at about 2.7\,runs/ms. The GPU has 240
processors, which can work in parallel. Naively one would therefore
expect the performance to improve until 240 inspirals are performed in
parallel and after that the performance to level off as the processors
need to work through the inspirals serially. However, the runtime CUDA
scheduler has to be able to keep these 240 processors working
simultaneously without waiting for memory access, which can be
extremely slow. This is the likely reason why the performance still
rises even after $N=10,000$, as the runtime has a better chance of
squeezing the optimal performance out of the card by interleaving
different inspirals without having to wait for costly memory
transfers. We performed this study using block sizes of 256
threads. We found similar behavior with 128-thread blocks.

Based on this data, we achieve a very impressive \emph{speed-up of about a factor 50}. While this comparison is slightly unfair to the
CPU because we only use a single core and double-precision on the CPU,
there is no doubt the performance gains are very significant. Note
also that for this test problem we integrated the exact same problem
in all cases. This means the individual threads run in perfect
lock-step, the best possible case for the GPU. We typically see
differences of about 10\% in the runtime of different inspirals, and
this would result in a slight inefficiency on the GPU as some of the
threads in a block may finish before others. The double versus
single-precision on the CPU has only a moderate effect, and the
parallelization on the CPU will only be simple as long as the problem
remains of the ``embarrassingly parallel'' type. We plan to extend our
studies toward more dynamic explorations of phase space requiring
dynamic feedback and communication between individual results. This
would then require a significant implementation effort on the CPU in
actually parallelizing the code, which is likely to be more difficult
than the conversion to CUDA.

\section{Some Inspiral Results}

In this section we highlight some of our findings from a completed
initial study of the inspiral space. We emphasize that at this point
we have focused on a few select situations only. We plan to soon move
on to study larger regions of the initial configuration space and use more advanced
statistical analysis tools to try to dynamically identify
``interesting'' regions. 

For these simulations we integrate the ODEs
Eq.~(\ref{eq:domdt})-(\ref{eq:dLdt}) in the range $\omega_0=0.00395
\rightarrow \omega_f=0.05$. The initial
orbital frequency $\omega_0$ corresponds to a separation of $r\approx
40 M$, and the final $\omega_f$ is chosen such that the gravitational
wave frequency still matches numerical relativity results. We have
varied both frequencies and, except in one case, have not found significant
qualitative differences in the results. We leave detailed experimentation
with different $\omega_0$ and $\omega_f$ as a topic of future study.
The initial unit angular momentum
is chosen as the unit vector in the z-direction
$\hat{\mathbf{L}}_n=(0,0,1)$ without loss of generality.

For each choice of the black holes masses and spin magnitudes 
we sample in a uniform and random way their spin {\em orientations}. The latter corresponds 
to uniformly sampling a sphere, using the algorithm
of~\cite{Marsaglia:1972}. We use $N=100,000$ randomly selected
initial spin orientations for each black hole. That is, a total of $N=10^8$ spin configurations.

In analyzing the data we have found the scalar products between the
different unit vectors a useful quantity to investigate. This is also
motivated by the crucial role these scalar products play in the recoil
velocity.

\subsection{Equal-mass, maximally spinning black
  holes}\label{sec:eq_m_max_chi}

We start by looking at the case of two equal-mass ($m_i=0.5$), maximally
spinning ($\chi_i=1$) black holes  with randomly oriented initial
unit-vector spin configurations. For this case all the pre-factors in
the ODEs Eq.~(\ref{eq:domdt})-Eq.~(\ref{eq:dLdt}) for the spin terms
are identical and the $\sum_i \chi_i
\hat{\mathbf{L}}\cdot\hat{\mathbf{S}}_i$ term in the RHS for $\omega$ in
particular simplifies to a form $\hat{\mathbf{L}}\cdot\mathbf{S}$ with
$\mathbf{S}=\mathbf{S_1}+\mathbf{S_2}$. Note however that there is
still a term of the form $\hat{\mathbf{S}}_1\cdot\hat{\mathbf{S}}_2$,
which will change for the different configurations used.

\begin{figure}
\includegraphics[width=0.4\textwidth]{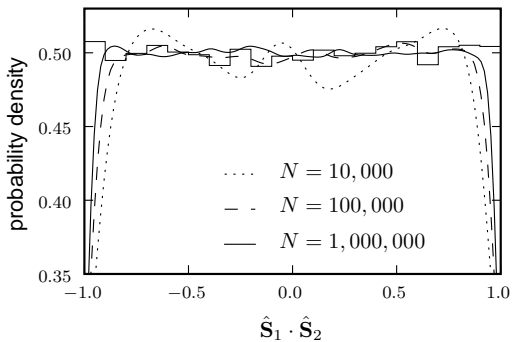}
\caption{Uniform distribution of the scalar product
  $\hat{\mathbf{S}}_1\cdot\hat{\mathbf{S}}_2$ of the final
  configuration at $\omega_f$. The plot shows the convergence of the
  Gaussian kernel estimated probability density function for different $N$ towards the uniform
  distribution case (which would be at 0.5) as well as a histogram for
  the $N=100,000$ case.}
\label{fig:kpdf}
\end{figure}

As a first measure of the spin dynamics, we look at the scalar product
between the two unit spin vectors
$\hat{\mathbf{S}}_1\cdot\hat{\mathbf{S}}_2$ as well as the scalar
products with the unit angular momentum
$\hat{\mathbf{S}}_i\cdot\hat{\mathbf{L}}_n$. The simulations start out
from an initially uniform distribution in each of these scalar
products. Fig.~\ref{fig:kpdf} shows a histogram of the final value of
$\hat{\mathbf{S}}_1\cdot\hat{\mathbf{S}}_2$ for $N=100,000$ inspirals
and $N_{\text{bins}}=20$ bins. The figure also shows the 
probability density estimator for the final value of
$\hat{\mathbf{S}}_1\cdot\hat{\mathbf{S}}_2$ at $\omega_f$, using 
 a Gaussian kernel for different choices of the number of
inspirals $N$. The density estimator is given by
$\hat{p}(x)=1/(Nh) \sum_j^N K\left((x-x_j)/h\right)$, where $h$ is the
bandwidth and for the Gaussian kernel:
$K\left((x-x_i)/h\right)=1/\sqrt{2\pi}\exp\left(-(x-x_i)^2/2h^2\right)$. We
construct the bandwidth $h$ using $h=1.06\sigma N^{-1/5}$, where $\sigma$ is the standard deviation and
$N$ is the number of inspirals~\cite{Silverman86}. As $N$ is increased,
the probability density estimator goes toward a uniform
distribution. This shows that there is no structure in the
final spin scalar product, i.e. that it is uniformly distributed. Note
that the steep fall-off near the boundaries is an artifact of the
estimator, which assumes the variable is distributed in the real line 
instead of $[-1,1]$ and smoothes out the discontinuous jump at $\pm
1$. This feature converges away in $N$. For
$\hat{\mathbf{S}}_i\cdot\hat{\mathbf{L}}_n$ the distributions look
very similar. In the equal-mass, maximum spin case there is no
preferred angle between the spin vectors generated in the inspiral.

\begin{figure}
\includegraphics[width=0.4\textwidth]{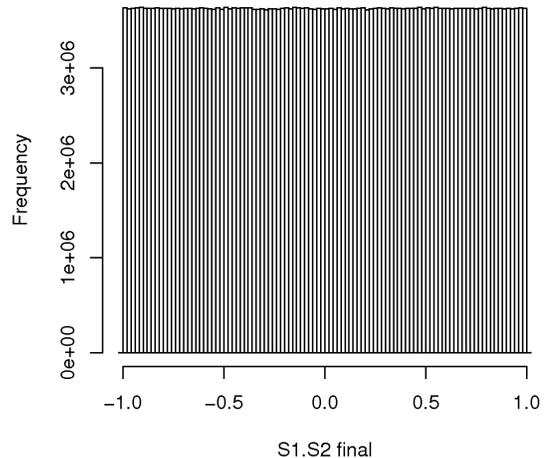}
\caption{Uniform distribution of the scalar product
  $\hat{\mathbf{S}}_1\cdot\hat{\mathbf{S}}_2$ of the final
  configuration at $\omega_f$ when BH masses are even and
  $\chi_i=0.7$. The plot shows a histogram for the $N=362,799,815$ case.}
\label{fig:hugehisto}
\end{figure}

To further validate this even distribution of final dot product values,
we used the Kolmogorov-Smirnov test to measure the uniformness of the final
values. For this test, the BHs had equal mass and were non-maximally
spinning, with $\chi_i=0.7$. We used a large sample containing 362,799,815
inspirals with random $\hat{\mathbf{S}}_i$ vectors over the unit sphere. 
The K-S test
returned a p-value of $5.17\times 10^{-5}$, indicating a very uniform final
distribution. The final histogram for this large test is displayed in
Fig.~\ref{fig:hugehisto}, using $N_{\text{bins}}=100$ bins. It is visually
apparent that no final dot product of the spin vectors is favored.

\begin{figure}
\includegraphics[width=0.4\textwidth]{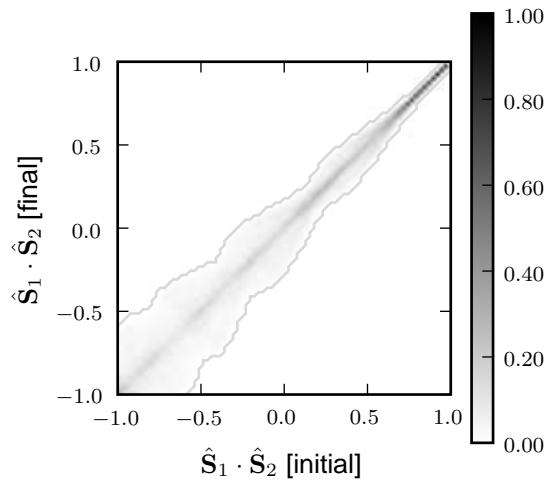}
\caption{Final versus initial spin scalar product
  $\hat{\mathbf{S}}_1\cdot\hat{\mathbf{S}}_2$. The plot shows a high
  correlation in the scalar product between initial and final scalar
  product for the equal mass ($m_1=m_2=0.5$), maximally spinning
  ($\chi_1=\chi_2=1$) case. In light gray we show the 0-contour line
  to make it easier to see how far the configurations have spread.}
\label{fig:s1s2corr_em_ms}
\end{figure}

\begin{figure}
\includegraphics[width=0.4\textwidth]{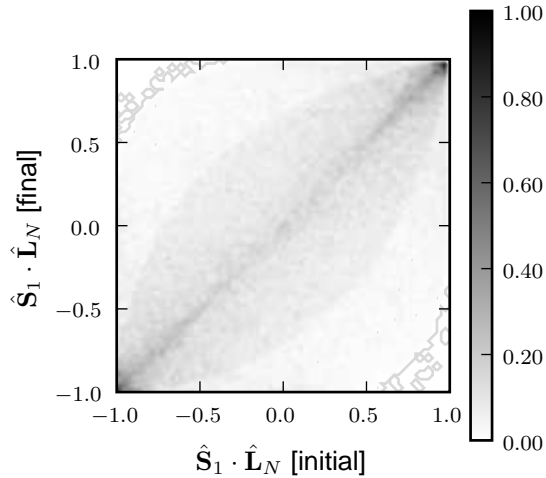}
\caption{Final versus initial scalar product 
  $\hat{\mathbf{S}}_1\cdot\hat{\mathbf{L}}_n$ for
  the same case as in Fig.~\ref{fig:s1s2corr_em_ms}. There is much less correlation between
  these initial and final scalar products compared to  $\hat{\mathbf{S}}_1\cdot \hat{\mathbf{S}}_2$.}
\label{fig:s1lncorr_em_ms}
\end{figure}

While the final distribution in
$\hat{\mathbf{S}}_1\cdot\hat{\mathbf{S}}_2$ remains uniform, we do find
a high linear correlation between the initial and final values of this scalar product,
as can be seen in Fig.~\ref{fig:s1s2corr_em_ms}. The latter shows a bivariate histogram
for  $(\hat{\mathbf{S}}_1\cdot\hat{\mathbf{S}}_2)_i$ and  $(\hat{\mathbf{S}}_1\cdot\hat{\mathbf{S}}_2)_f$. Note that the
histogram is normalized such that the maximum value is 1. We have also
plotted the 0-contour line in light gray to make it easier to see how
far configurations have spread. We have used $N_{\text{hist}}=79$ bins
in each dimension, which was found by following the procedure in
Ref.~\cite{2008arXiv0807.4820H} and forcing equal numbers of bins in
both dimensions.
This linear correlation can be quantified using the Pearson product-moment
coefficient which is given by $r=1/(N-1) \sum_{i=1}^N
\left((x_i-\bar{x})/\sigma_x\right)\left((y_i-\bar{y})/\sigma_y\right)$, 
where $\bar{x}, \bar{y}$ are the sample means, $\sigma_x, \sigma_y$
are the standard deviations, and $N$ is the number of inspirals. For
the case of Fig.~\ref{fig:s1s2corr_em_ms} we find $r=0.99$. Note that
the correlations for $\hat{\mathbf{S}}_1\cdot\hat{\mathbf{L}}_n$ and
$\hat{\mathbf{S}}_2\cdot\hat{\mathbf{L}}_n$ are not nearly as high,
with $r=0.65$ for both cases. In Fig.~\ref{fig:s1lncorr_em_ms} we show
a similar histogram plot for
$\hat{\mathbf{S}}_1\cdot\hat{\mathbf{L}}_n$. The case for
$\hat{\mathbf{S}}_2\cdot\hat{\mathbf{L}}_n$ looks similar. The lower
correlation is clearly visible compared to
Fig.~\ref{fig:s1s2corr_em_ms}, as the values are much further spread
from the diagonal in Fig.~\ref{fig:s1lncorr_em_ms}.

\subsection{Unequal-mass, maximally spinning black
  holes}\label{sec:uneq_m_max_chi}
\begin{figure}
\includegraphics[width=0.4\textwidth]{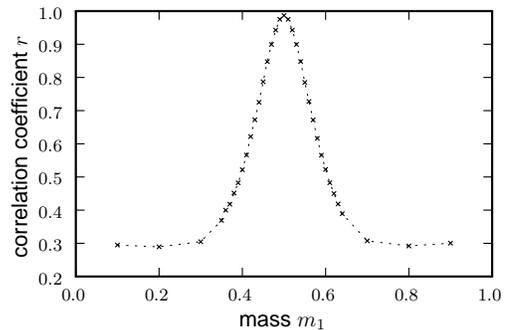}
\caption{The correlation coefficient $r$ of
  $(\hat{\mathbf{S}}_1\cdot\hat{\mathbf{S}}_2)_i$ and
  $(\hat{\mathbf{S}}_1\cdot\hat{\mathbf{S}}_2)_f$ for maximally
  spinning black holes as the mass of the bodies is changed. The
  correlation falls quickly as the masses become uneven.}
\label{fig:pears_mass}
\end{figure}

Having found the high correlation between the initial and final scalar products of
 spin vectors mentioned above we now change the mass
$m_1$ of one BH while keeping $\chi_1=\chi_2=1$ and study the
resulting change in correlation. In Fig.~\ref{fig:pears_mass} we show
the correlation coefficient $r$ of the initial and final values of
$\hat{\mathbf{S}}_1\cdot\hat{\mathbf{S}}_2$ as the mass $m_1$
changes. Recall that we normalized the masses such that $M=m_1+m_2=1$,
and hence $m_1\in [0,1]$.
Fig.~\ref{fig:pears_mass} shows that the correlation drops quickly as
the masses become uneven. This would indicate it would be harder to
predict in a probabilistic sense the final value of this scalar product for a  
of a BBH inspiral in which the BHs have dissimilar masses and high spins. 

\subsection{Equal-mass, equal-spin, non-maximally spinning}
\begin{figure}
\includegraphics[width=0.4\textwidth]{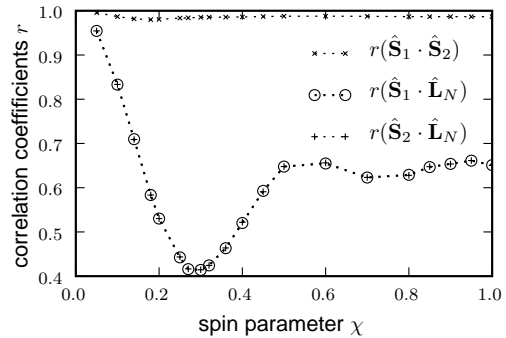}
\caption{The correlation coefficient $r$ between the initial and final 
  values of $\hat{\mathbf{S}}_1\cdot\hat{\mathbf{S}}_2$ and
  $\hat{\mathbf{S}}_j\cdot\hat{\mathbf{L}}_n$ ($j=1,2$). The masses are kept  constant and equal
  ($m_1=m_2=0.5$) while the spin parameters are changed at the same rate, $\chi_1= \chi_2 = \chi$. The
  correlation remains nearly constant and large for the
  $\hat{\mathbf{S}}_1\cdot\hat{\mathbf{S}}_2$ scalar product, but varies for
  $\hat{\mathbf{S}}_j\cdot\hat{\mathbf{L}}_n$.}
\label{fig:pears_chi}
\end{figure}

Next we vary the spin $\chi$ for each black hole while keeping the same rate ($\chi_1=\chi_2$), 
and fixing the masses to be even ($m_1=m_2=0.5$) to
see if there is a similar behavior in the correlation between
$(\hat{\mathbf{S}}_1\cdot\hat{\mathbf{S}}_2)_i$ and
$(\hat{\mathbf{S}}_1\cdot\hat{\mathbf{S}}_2)_f$ as we found when changing 
$m_1$. A similar behavior could be expected as the mass and
spin parameters enter similarly in the spin evolution
equations. However, as Fig.~\ref{fig:pears_chi} shows, the correlation
coefficient $r$ of the initial and final values of
$\hat{\mathbf{S}}_1\cdot\hat{\mathbf{S}}_2$ remains constantly large
($r>0.98$), unlike the case of changing $m_1$ above. In
Fig.~\ref{fig:pears_chi} we also show the correlation coefficient $r$
for $\hat{\mathbf{S}}_j\cdot\hat{\mathbf{L}}_n$ ($j=1,2$), which shows
significant variation.
In the case of $\hat{\mathbf{S}}_j\cdot\hat{\mathbf{L}}_n$ the
behavior is similar for each $j$, which is expected since both BHs enter
symmetrically into the expressions. The only difference between the
BHs is their initial location.

\subsection{$m_1=0.4$, equal-spin $\chi=0.05$}
\begin{figure}
\includegraphics[width=0.4\textwidth]{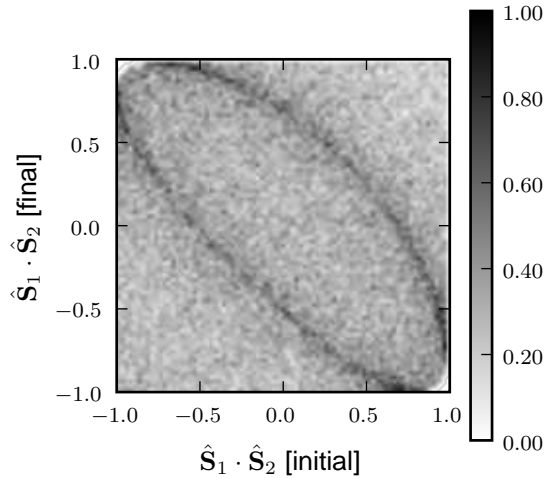}
\caption{Final versus initial scalar product of spin vectors
  $\hat{\mathbf{S}}_1\cdot\hat{\mathbf{S}}_2$ for $m_1=0.4$ and
  equal-spin $\chi=0.05$ as an example of the richer structure away
  from highly symmetric cases. Note that the correlation is now
  negative.}
\label{fig:corr_ring}
\end{figure}

As an example of the rich structure that can be found if one leaves
the equal mass case, we show in Fig.~\ref{fig:corr_ring} a plot of the
final versus initial spin scalar product
$\hat{\mathbf{S}}_1\cdot\hat{\mathbf{S}}_2$ for the case of $m_1=0.4$,
and $\chi_1=\chi_2=\chi=0.05$. In this case a ring-like structure
appears in the scalar product and the data is now anti-correlated with
$r=-0.14$. Though in other investigations presented in this paper the
results were generally insensitive to the values of $\omega_0$ and $\omega_f$,
we found in this case that the structure of the bivariate histogram
could vary significantly based on these values. We leave a more full
investigation of these cases to future work.

\section{Discussion}
We have described our implementation of parallel integrations of the
PN equations describing the circular inspiral of 2 black holes. Using
the inherent parallelism of the problem of the inspiral space study of
this system, we chose to implement this on Graphics Processors well
adapted to the task. We achieved a speed-up of $50\times$ compared to
a single-core CPU. This speed-up will be important for more advanced
studies of large numbers of inspirals planned for the future.

First results from initial studies indicate there is a rich structure
for certain regions of initial inspiral conditions. For an example see
Fig.~\ref{fig:corr_ring}. In particular, the equal-mass case appears
special, leading to more correlated dynamics compared to the
equal-spin case. We used scalar products between the unit vectors to
judge the dynamics.

For the future we plan to significantly expand our studies of the
initial configuration space and to use more advanced statistical methods for analysis.

\acknowledgments This work was supported in part by the NSF grant
PHY0801213 to the University of Maryland and an NVIDIA professor partnership. 
We would like to thank Yi Pan and Alessandra Buonanno for multiple discussions, 
William Dorland, Ramani Duraiswami, Nail Gumerov and George Stantchev for introducing us 
to GPU computing and Saul Teukolsky for pointing us to~\cite{Marsaglia:1972}. The 
simulations of this paper were done using two NVIDIA Tesla S1070 Computing Systems.

\bibliographystyle{unsrt}
\bibliography{references}

\end{document}